\documentstyle[prl,aps,multicol,eqsecnum]{revtex}

\draft

\author{Isaac L. Chuang $^{1,2}$ {\em and}\/ M. A. Nielsen $^{1,3}$}

\title{Prescription for experimental determination of \\ the dynamics of
	a quantum black box} 

\address{\vspace*{1.2ex}
 	\hspace*{0.5ex}{$^1$ Institute for Theoretical Physics \\
	University of California, Santa Barbara, CA 93106-4030 }}
\address{\vspace*{1.2ex}
 	\hspace*{0.5ex}{$^2$ ERATO Quantum Fluctuation Project \\
 	Edward L. Ginzton Laboratory, Stanford University, 
 		Stanford, CA 94305-4085}}
\address{\vspace*{1.2ex}
 	\hspace*{0.5ex}{$^3$ Center for Advanced Studies, Department
	of Physics and Astronomy, \\
	University of New Mexico, NM 87131-1156}}

\date{\today}

\begin{document}

\pagestyle{plain}
\pagenumbering{arabic}

\maketitle


\begin{abstract}
We give an explicit prescription for experimentally determining the evolution
operators which completely describe the dynamics of a quantum mechanical
black box -- an arbitrary open quantum system.  We show necessary and
sufficient conditions for this to be possible, and illustrate the general
theory by considering specifically one and two quantum bit systems.  These
procedures may be useful in the comparative evaluation of experimental
quantum measurement, communication, and computation systems.
\end{abstract}



\begin{multicols}{2}[]

\def\be{\begin{equation}}
\def\ee{\end{equation}}
\def\bea{\begin{eqnarray}}
\def\eea{\end{eqnarray}}
\def\Adag{A^\dagger}
\newcommand{\mattwoc}[4]{\left[
	\begin{array}{cc}{#1}&{#2}\\{#3}&{#4}\end{array}\right]}
\newcommand{\ket}[1]{\mbox{$|#1\rangle$}}
\newcommand{\bra}[1]{\mbox{$\langle #1|$}}
\def\>{\rangle}
\def\<{\langle}

\section{Introduction}

Consider a black box with an input and an output.  Given that the transfer
function is linear, if the dynamics of the box are described by classical
physics, well known recipes exist to completely determine the response
function of the system.  Now consider a {\em quantum-mechanical} black box
whose input may be an arbitrary quantum state (in a finite dimensional Hilbert
space), with internal dynamics and an output state (of same dimension as the
input) determined by quantum physics.  The box may even be connected to an
external reservoir, or have other inputs and outputs which we wish to ignore.
Can we determine the quantum transfer function of the system?

The answer is yes.  Simply stated, the most arbitrary transfer function of a
quantum black box is to map one density matrix into another, $\rho_{in}
{\rightarrow} \rho_{out}$, and this is determined by a linear mapping ${\cal
E}$ which we shall give a prescription for obtaining.  The interesting
observation is that this black box may be an attempt to realize a useful
quantum device.  For example, it may be a quantum cryptography
channel\cite{Bennett92,Hughes95} (which might include an eavesdropper!), a
quantum computer in which decoherence occurs, limiting its
performance\cite{Unruh94,Chuang95a}, or just an imperfect quantum logic
gate\cite{Turchette95,Monroe95}, whose performance you wish to characterize to
determine its usefulness.

How many parameters are necessary to describe a quantum black box acting on
an input with a state space of $N$ dimensions?  And how may these parameters
be experimentally determined?  Furthermore, how is the resulting description
of ${\cal E}$ useful as a performance characterization?

We consider these questions in this paper.  After summarizing the relevant
mathematical formalism, we prove that ${\cal E}$ may be determined completely
by a matrix of complex numbers $\chi$, and provide an accessible experimental
prescription for obtaining $\chi$.  We then give explicit constructions for
the cases of one and two quantum bits (qubits), and then conclude by
describing related performance estimation quantities derivable from $\chi$.

\section{State Change Theory}

A general way to describe the state change experienced by a quantum system is
by using {\em quantum operations}, sometimes also known as {\em
superscattering operators} or {\em completely positive maps}. This formalism
is described in detail in \cite{Kraus83a}, and is given a brief but
informative review in the appendix to \cite{Schumacher96a}. A quantum
operation is a linear map ${\cal E}$ which completely describes the dynamics
of a quantum system,
\begin{equation}
	\rho \rightarrow \frac{{\cal E}(\rho)}{\mbox{tr}({\cal E}(\rho))}
\,.
\label{eq:rhomapfirst}
\end{equation}
A particularly useful description of quantum operations for theoretical
applications is the so-called {\em operator-sum representation}:
\begin{equation} \label{eqtn: op sum rep}
	{\cal E}(\rho) = \sum_i A_i \rho A_i^{\dagger}
\,.
\label{eq:eeffect}
\end{equation}
The $A_i$ are operators acting on the system alone, yet they completely
describe the state changes of the system, including any possible unitary
operation (quantum logic gate), projection (generalized measurement), or
environmental effect (decoherence).  In the case of a ``non-selective''
quantum evolution, such as arises from uncontrolled interactions with an
environment (as in the decoherence of quantum computers), the $A_i$ operators
satisfy an additional completeness relation,
\begin{eqnarray}
	\sum_i A_i^{\dagger} A_i = I
\,. 
\label{eq:completeness}
\end{eqnarray}
This relation ensures that the trace factor $\mbox{tr}({\cal E}(\rho))$ is
always equal to one, and thus the state change experienced by the system can
be written
\begin{equation}
	\rho \rightarrow {\cal E}(\rho)
\,. 
\label{eq:rhomap}
\end{equation}

Such quantum operations are in a one to one correspondence with the set of
transformations arising from the joint unitary evolution of the quantum system
and an initially uncorrelated environment\cite{Kraus83a}.  In other words, the
quantum operations formalism also describes the master equation and quantum
Langevin pictures widely used in quantum optics \cite{Louisell,Gardiner91},
where the system's state change arises from an interaction Hamiltonian between
the system and its environment\cite{Mabuchi96}.

Our goal will be to describe the state change process by determining the
operators $A_i$ which describe ${\cal E}$, (and until Section~\ref{sec:meas}
we shall limit ourselves to those which satisfy Eq.(\ref{eq:completeness})).
Once these operators have been determined many other quantities of great
interest, such as the {\em fidelity}, {\em entanglement fidelity} and {\em
quantum channel capacity} can be determined.  Typically, the $A_i$ operators
are derived from a {\em theoretical} model of the system and its environment;
for example, they are closely related to the Lindblad operators.  However,
what we propose here is different: to determine systematically from {\em
experiment} what the $A_i$ operators are for a specific quantum black box.

\section{General Experimental Procedure}

The experimental procedure may be outlined as follows.  Suppose the state
space of the system has $N$ dimensions; for example, $N=2$ for a single qubit.
$N^2$ pure quantum states $|\psi_1\>\<\psi_1|,
\ldots,|\psi_{N^2}\>\<\psi_{N^2}|$ are experimentally prepared, and the output
state ${\cal E}(|\psi_j\>\<\psi_j|)$ is measured for each input.  This may be
done, for example, by using quantum state
tomography\cite{Raymer94a,Leonhardt96,Leibfried96a}.  In principle, the
quantum operation ${\cal E}$ can now be determined by a linear extension of
${\cal E}$ to all states.  We prove this below.

The goal is to determine the unknown operators $A_i$ in Eq.(\ref{eq:eeffect}).
However, experimental results involve numbers (not operators, which are a
theoretical concept).  To relate the $A_i$ to measurable parameters, it is
convenient to consider an equivalent description of ${\cal E}$ using a {\em
fixed} set of operators $\tilde{A}_i$, which form a basis for the set of
operators on the state space, so that
\begin{eqnarray}
		A_i = \sum_m a_{im} \tilde{A}_m
\label{eq:atildedef}
\end{eqnarray}
for some set of complex numbers $a_{im}$.  Eq.(\ref{eq:eeffect}) may
thus be rewritten as
\be 
\label{eqtn: two sided rep}
	{\cal E}(\rho) = \sum_{mn} \tilde{A}_m \rho 
			\tilde{A}_{n}^{\dagger} \chi_{mn}
\,,
\ee
where $\chi_{mn} \equiv \sum_i a_{im} a_{in}^*$ is a ``classical'' {\em error
correlation matrix} which is positive Hermitian by definition.  This shows
that ${\cal E}$ can be completely described by a complex number matrix,
$\chi$, once the set of operators $\tilde{A}_i$ has been fixed.  In general,
$\chi$ will contain $N^4-N^2$ independent parameters, because a general linear
map of $N$ by $N$ matrices to $N$ by $N$ matrices is described by $N^4$
independent parameters, but there are $N^2$ additional constraints due to the
fact that the trace of $\rho$ remains one. We will show how to determine
$\chi$ experimentally, and then show how an operator sum representation of the
form Eq.(\ref{eqtn: op sum rep}) can be recovered once the $\chi$ matrix is
known.

Let $\rho_j$, $1\leq j \leq N^2$ be a set of linearly independent basis
elements for the space of $N$$\times$$N$ matrices.  A convenient choice is the
set of projectors $\ket{n}\bra{m}$.  Experimentally, the output state ${\cal
E}(\ket{n}\bra{m})$ may be obtained by preparing the input states $\ket{n}$,
$\ket{m}$, $\ket{n_+} = (\ket{n}+\ket{m})/\sqrt{2}$, and $\ket{n_-} =
(\ket{n}+i\ket{m})/\sqrt{2}$ and forming linear combinations of ${\cal
E}(\ket{n}\bra{n})$, ${\cal E}(\ket{m}\bra{m})$, ${\cal
E}(\ket{n_+}\bra{n_+})$, and ${\cal E}(\ket{n_-}\bra{n_-})$.  Thus, it is
possible to determine ${\cal E}(\rho_j)$ by state tomography, for each
$\rho_j$.

Furthermore, each ${\cal E}(\rho_j)$ may be expressed as a linear combination
of the basis states,
\be
	{\cal E}(\rho_j)
	= \sum_k \lambda_{jk} \rho_k
\,,
\ee
and since ${\cal E}(\rho_j)$ is known, $\lambda_{jk}$ can thus be determined.
To proceed, we may write
\be
	\tilde{A}_m \rho_j \tilde{A}_n^\dagger = \sum_k \beta^{mn}_{jk} \rho_k
\,,
\label{eq:betadef}
\ee
where $\beta^{mn}_{jk}$ are complex numbers which can be determined by
standard algorithms given the $\tilde{A}_m$ operators and the $\rho_j$
operators.  Combining the last two expressions we have 
\be
	\sum_k \sum_{mn} \chi_{mn} \beta^{mn}_{jk} \rho_k 
	= \sum_k \lambda_{jk}\rho_k
\,.
\ee
{} From independence of the $\rho_k$ it follows that for each $k$,
\be
\label{eqtn: chi condition}
	\sum_{mn} \beta^{mn}_{jk} \chi_{mn} = \lambda_{jk}
\,.
\ee
This relation is a necessary and sufficient condition for the matrix
$\chi$ to give the correct quantum operation ${\cal E}$.  One may
think of $\chi$ and $\lambda$ as vectors, and $\beta$ as a
$N^4$$\times$$N^4$ matrix with columns indexed by $mn$, and rows by
$ij$.  To show how $\chi$ may be obtained, let $\kappa$ be the
generalized inverse for the matrix $\beta$, satisfying the relation
\begin{equation}
	\beta^{mn}_{jk} = \sum_{st,xy} \beta_{jk}^{st} \kappa_{st}^{xy}
		 \beta_{xy}^{mn}
\,.
\end{equation}
Most computer algebra packages are capable of finding such generalized
inverses.  In appendix \ref{appendix: chi} it is shown that $\chi$
defined by
\begin{eqnarray}
	\chi_{mn} = \sum_{jk} \kappa_{jk}^{mn} \lambda_{jk}
\label{eqtn:chidefn}
\end{eqnarray}
satisfies the relation (\ref{eqtn: chi condition}). The proof is
somewhat subtle, but it is not relevant to the application of the
present algorithm.

Having determined $\chi$ one immediately obtains the operator sum
representation for ${\cal E}$ in the following manner.  Let the unitary matrix
$U^\dagger$ diagonalize $\chi$,
\begin{eqnarray}
	\chi_{mn} = \sum_{xy} U_{mx} d_{x} \delta_{xy} U^*_{ny}
. 
\end{eqnarray}
{} From this it can easily be verified that
\begin{eqnarray}
	A_i = \sqrt{d_i} \sum_j U_{ij} \tilde{A}_j
\end{eqnarray}
gives an operator-sum representation for the quantum operation ${\cal
E}$.  Our algorithm may thus be summarized as follows: $\lambda$ is
experimentally measured, and given $\beta$, determined by a choice of
$\tilde{A}$, we find the desired parameters $\chi$ which completely
describe ${\cal E}$.

\section{One and Two Qubits}

The above general method may be illustrated by the specific case of a
black box operation on a single quantum bit (qubit).  A convenient
choice for the fixed operators $\tilde{A}_i$ is
\bea
	\tilde{A}_0 &=& I
\label{eq:fixedonebit}
\\	\tilde{A}_1 &=& \sigma_x
\\	\tilde{A}_2 &=& -i \sigma_y
\\	\tilde{A}_3 &=& \sigma_z
\,,
\label{eq:fixedonebitend}
\eea
where the $\sigma_i$ are the Pauli matrices.  There are 12 parameters,
specified by $\chi$, which determine an arbitrary single qubit black box
operation ${\cal E}$; three of these describe arbitrary unitary transforms
$\exp(i\sum_k r_k\sigma_k)$ on the qubit, and nine parameters describe
possible correlations established with the environment $E$ via
$\exp(i\sum_{jk} \gamma_{jk} \sigma_j\otimes\sigma^E_k)$.  Two combinations of
the nine parameters describe physical processes analogous to the $T_1$ and
$T_2$ spin-spin and spin-lattice relaxation rates familiar to us from
classical magnetic spin systems.  However, the dephasing and energy loss rates
determined by $\chi$ do not simply describe ensemble behavior; rather, $\chi$
describes the dynamics of a {\em single quantum system}.  Thus, the
decoherence of a single qubit must be described by {\em more than just two
parameters}.  {\em Twelve} are needed in general.

These 12 parameters may be measured using four sets of experiments.  As a
specific example, suppose the input states $\ket{0}$, $\ket{1}$,
$\ket{+}=(\ket{0}+\ket{1})/\sqrt{2}$ and $\ket{-} =
(\ket{0}+i\,\ket{1})/\sqrt{2}$ are prepared, and the four matrices
\bea
	\rho'_1 &=& {\cal E}(\ket{0}\bra{0})
\\	\rho'_4 &=& {\cal E}(\ket{1}\bra{1})
\\	\rho'_2 &=& {\cal E}(\ket{+}\bra{+}) 
			- i {\cal E}(\ket{-}\bra{-})
			- (1-i)(\rho'_1 + \rho'_4)/2
\\	\rho'_3 &=& {\cal E}(\ket{+}\bra{+}) 
			+ i {\cal E}(\ket{-}\bra{-})
			- (1+i)(\rho'_1 + \rho'_4)/2
\eea
are determined using state tomography.  These correspond to $\rho'_j = {\cal
E}(\rho_j)$, where
\be
	\rho_1 = \mattwoc{1}{0}{0}{0}
\,,
\ee
$\rho_2 = \rho_1 \sigma_x$, $\rho_3=\sigma_x\rho_2$, and $\rho_4 = \sigma_x
\rho_1\sigma_x$.  From Eq.(\ref{eq:betadef}) and
Eqs.(\ref{eq:fixedonebit}-\ref{eq:fixedonebitend}) we may determine $\beta$,
and similarly $\rho'_j$ determines $\lambda$.  However, due to the particular
choice of basis, and the Pauli matrix representation of $\tilde{A}_i$, we may
express the $\beta$ matrix as the Kronecker product $\beta = \Lambda\otimes
\Lambda$, where
\be
	\Lambda = \frac{1}{2} \mattwoc{I}{\sigma_x}{\sigma_x}{-I}
\,,
\ee
so that $\chi$ may be expressed conveniently as
\be
	\chi = \Lambda \mattwoc{\rho'_1}{\rho'_2}{\rho'_3}{\rho'_4} \Lambda
\,,
\ee
in terms of block matrices.  

%
%
%
%

Likewise, it turns out that the parameters $\chi_2$ describing the
black box operations on two qubits can be expressed as
\be
	\chi_2 = \Lambda_2 \overline{\rho}' \Lambda_2
\,,
\ee
where $\Lambda_2 = \Lambda \otimes \Lambda$, and $\overline{\rho}'$ is
a matrix of sixteen measured density matrices,
\be
	\overline{\rho}' = P^T
	\left[\begin{array}{cccc}
		\rho'_{11} & \rho'_{12} & \rho'_{13} & \rho'_{14}
	\\	\rho'_{21} & \rho'_{22} & \rho'_{23} & \rho'_{24}
	\\	\rho'_{31} & \rho'_{32} & \rho'_{33} & \rho'_{34}
	\\	\rho'_{41} & \rho'_{42} & \rho'_{43} & \rho'_{44}
	\end{array}\right]
	P
\,,
\ee
where $\rho'_{nm} = {\cal E}(\rho_{nm})$, $\rho_{nm} = T_n
\ket{00}\bra{00} T_m$, $T_1 = I\otimes I$, $T_2 = I\otimes \sigma_x$,
$T_3 = \sigma_x \otimes I$, $T_4 = \sigma_x \otimes \sigma_x$, and $P
= I\otimes [(\rho_{00}+\rho_{12}+\rho_{21}+\rho_{33})\otimes I]$ is a
permutation matrix.  Similar results hold for $k>2$ qubits.  Note that
in general, a quantum black box acting on $k$ qubits is described by
$16^k-4^k$ independent parameters.

There is a particularly elegant geometric view of quantum
operations for a single qubit. This is based on the Bloch vector,
$\vec \lambda$, which is defined by
\begin{equation}
	\rho = \frac{I+\vec \lambda \cdot \vec \sigma}{2},
\end{equation}
satisfying $| \vec \lambda | \leq 1$. The map Eq.(\ref{eq:rhomap})
is equivalent to a map of the form
\begin{equation}
	\vec \lambda \stackrel{\cal E}{\rightarrow} \vec \lambda' 
		= M \vec \lambda + \vec c
\,,
\label{eqtn: affine map}
\end{equation}
where $M$ is a $3$$\times$$3$ matrix, and $\vec c$ is a constant
vector. This is an {\em affine map}, mapping the Bloch sphere into
itself. If the $A_i$ operators are written in the form
\begin{eqnarray}
	A_i = \alpha_i I + \sum_{k=1}^3 a_{ik} \sigma_k,
\end{eqnarray}
then it is not difficult to check that
\begin{eqnarray}
	M_{jk} & = &  \sum_l \left[ \begin{array}{l}
		a_{lj} a_{lk}^* + a_{lj}^* a_{lk}
	+  \\ \left( |\alpha_l|^2- \sum_p a_{lp} a_{lp}^* \right) \delta_{jk}   
	+  \\
	i \sum_p \epsilon_{jkp}
	    ( \alpha_l a_{lp}^* - \alpha_l^* a_{lp} ) 
	\end{array} \right]
\\
	c_k &=& 2i \sum_l \sum_{jp} \epsilon_{jpk} a_{lj} a_{lp}^*
\,,
\end{eqnarray}
where we have made use of Eq.(\ref{eq:completeness}) to simplify the
expression for $\vec c$.

The meaning of the affine map Eq.(\ref{eqtn: affine map}) is made clearer
by considering the polar decomposition \cite{Horn91a} of the matrix $M$.
Any real matrix $M$ can always be written in the form
\begin{eqnarray}
	M = O S
\,, 
\end{eqnarray} 
where $O$ is a real orthogonal matrix with determinant $1$,
representing a proper rotation, and $S$ is a real symmetric
matrix. Viewed this way, the map Eq.(\ref{eqtn: affine map}) is just a
deformation of the Bloch sphere along principal axes determined by
$S$, followed by a proper rotation due to $O$, followed by a
displacement due to $\vec c$.  Various well-known decoherence measures
can be identified from $M$ and $\vec c$; for example, $T_1$ and $T_2$
are related to the magnitude of $\vec c$ and the norm of $M$.  Other
measures are described in the following section.

\section{Related Quantities}

We have described how to determine an unknown quantum operation ${\cal E}$ by
systematically exploring the response to a complete set of states in the
system's Hilbert space.  Once the operators $A_i$ have been determined, many
other interesting quantities can be evaluated.  A quantity of particular
importance is the {\em entanglement fidelity} \cite{Schumacher96a,Nielsen96c}.
This quantity can be used to measure how closely the dynamics of the quantum
system under consideration approximates that of some ideal quantum system.

Suppose the target quantum operation is a unitary quantum operation,
${\cal U}(\rho) = U \rho U^{\dagger}$, and the actual quantum
operation implemented experimentally is ${\cal E}$.  The entanglement
fidelity can be defined as \cite{Nielsen96c}
\begin{eqnarray}
	F_e(\rho,{\cal U},{\cal E}) 
	& \equiv & \sum_i \left|
		 \mbox{tr}(U^{\dagger} A_i \rho) \right|^2
\\
	&=& \sum_{mn} \chi_{mn} \mbox{tr} (U^{\dagger} \tilde{A}_m \rho)
			\mbox{tr}(\rho \tilde{A}_n^{\dagger} U)
\,.
\end{eqnarray}
The second expression follows from the first by using Eq.(\ref{eq:atildedef}),
and shows that errors in the experimental determination of ${\cal E}$
(resulting from errors in preparation and measurement) propagate linearly to
errors in the estimation of entanglement fidelity.  The minimum value of $F_e$
over all possible states $\rho$ is a single parameter which describes how well
the experimental system implements the desired quantum logic gate.

One may also be interested in the minimum {\em fidelity} of the gate
operation. This is given by the expression,
\begin{eqnarray}
	F \equiv \min_{|\psi\rangle} \langle \psi | U^{\dagger} 
		{\cal E}(|\psi\rangle \langle \psi|) U |\psi \rangle,
\end{eqnarray}
where the minimum is over all pure states, $|\psi\rangle$.  As for the
entanglement fidelity, we may show that this quantity can be
determined robustly, because of its linear dependence on the
experimental errors.

Another quantity of interest is the {\em quantum channel capacity}, defined by
Lloyd \cite{Lloyd96a,Schumacher96b} as a measure of the amount of quantum
information that can be sent using a quantum communication channel, such as an
optical fiber. In terms of the parameters discussed in this paper,
\begin{eqnarray}
	C({\cal E})
	\equiv \max_{\rho} S({\cal E}(\rho)) - S_e(\rho,{\cal E})
\,,
\end{eqnarray}
where $S({\cal E}(\rho))$ is the von Neumann entropy of the density operator
${\cal E}(\rho)$, $S_e(\rho,{\cal E})$ is the {\em entropy exchange}
\cite{Schumacher96a}, and the maximization is over all density operators
$\rho$ which may be used as input to the channel.  It is a measure of the
amount of quantum information that can be sent reliably using a quantum
communications channel which is described by a quantum operation ${\cal E}$.

One final observation is that our procedure can in principle be used to
determine the form of the Lindblad operator, ${\cal L}$, used in Markovian
master equations of the form
\begin{eqnarray}
	\dot \rho = {\cal L}(\rho),
\end{eqnarray}
where for convenience time is measured in dimensionless units, to make ${\cal
L}$ dimensionless.  This result follows from the fact that Lindblad operators
${\cal L}$ are just the logarithms of quantum operations; that is, $\exp({\cal
L})$ is a quantum operation for any Lindblad operator, ${\cal L}$, and $\log
{\cal E}$ is a Lindblad operator for any quantum operation ${\cal E}$.  This
observation may be used in the future to experimentally determine the form of
the Lindblad operator for systems, but will not be explored further here.

\section{Quantum Measurements}
\label{sec:meas}

Quantum operations can also be used to describe measurements.  For each
measurement outcome, $i$, there is associated a quantum operation, ${\cal
E}_i$. The corresponding state change is given by
\begin{eqnarray}
	\rho \rightarrow \frac{{\cal E}_i(\rho)}{\mbox{tr}({\cal E}_i(\rho))}
\,,
\end{eqnarray}
where the probability of the measurement outcome occurring is $p_i =
\mbox{tr}({\cal E}_i(\rho))$.  Note that this mapping may be {\em nonlinear},
because of this renormalization factor.  

Despite the possible nonlinearity, the procedure we have described may be
adapted to evaluate the quantum operations describing a measurement.  To
determine ${\cal E}_i$ we proceed exactly as before, except now we must
perform the measurement a large enough number of times that the probability
$p_i$ can be reliably estimated, for example by using the frequency of
occurrence of outcome $i$. Next, $\rho'_j$ is determined using tomography,
allowing us to obtain
\begin{eqnarray}
	{\cal E}_i(\rho_j) = \mbox{tr}({\cal E}_i(\rho_j)) \rho'_j,
\end{eqnarray}
for each input $\rho_j$ which we prepare, since each term on the right hand
side is known. Now we proceed exactly as before to evaluate the quantum
operation ${\cal E}_i$.  This procedure may be useful, for example, in
evaluating the effectiveness of a quantum-nondemolition (QND)
measurement\cite{braginsky92}.

\section{Conclusion}

In this paper we have shown how the dynamics of a quantum system may be
experimentally determined using a systematic procedure. This elementary {\em
system identification} step \cite{Ljung87} opens the way for robust
experimental determination of a wide variety of interesting quantities.
Amongst those that may be of particular interest are the quantum channel
capacity, the fidelity, and the entanglement fidelity.  We expect these
results to be of great use in the experimental study of quantum computation,
quantum error correction, quantum cryptography, quantum coding and quantum
teleportation.

\section*{Acknowledgments}

We thank C.~M.~Caves, R.~Laflamme, Y.~Yamamoto, and W.~H.~Zurek for many
useful discussions about quantum information and quantum optics. This work was
supported in part by the Office of Naval Research (N00014-93-1-0116), the
Phillips Laboratory (F29601-95-0209), and the Army Research Office
(DAAH04-96-1-0299).  We thank the Institute for Theoretical Physics for its
hospitality and for the support of the National Science Foundation
(PHY94-07194).  ILC acknowledges financial support from the Fannie and John
Hertz Foundation, and MAN acknowledges financial support from the
Australian-American Educational Foundation (Fulbright Commission).

\appendix
\section{Proof of the $\chi$ relation}
\label{appendix: chi}

The difficulty in verifying that $\chi$ defined by (\ref{eqtn:chidefn})
satisfies (\ref{eqtn: chi condition}) is that in general $\chi$
is not uniquely determined by the last set of equations. For
convenience we will rewrite these equations in matrix form as
\begin{eqnarray}
\label{eqtn: chi cond app}
	\beta \vec \chi & = & \vec \lambda \\
\label{eqtn: chi defn app}
	\vec \chi & \equiv & \kappa \vec \lambda
\,.
\end{eqnarray}
{} From the construction that led to equation (\ref{eqtn: two sided rep})
we know there exists at least one solution to equation
(\ref{eqtn: chi cond app}), which we shall call $\vec \chi '$. Thus
$\vec \lambda = \beta \vec \chi '$. The generalized inverse satisfies
$\beta \kappa \beta = \beta$.  Premultiplying the definition of $\vec \chi$
by $\beta$ gives
\begin{eqnarray}
	\beta \vec \chi & = & \beta \kappa \vec \lambda \\
	 & = & \beta \kappa \beta \vec \chi ' \\
	 & = & \beta \vec \chi ' \\
	 & = & \lambda 
\,.
\end{eqnarray}
Thus $\chi$ defined by (\ref{eqtn: chi defn app}) satisfies the equation
(\ref{eqtn: chi cond app}), as was required to show.

 

\begin{thebibliography}{10}

\bibitem{Bennett92}
C.~H. Bennett, G. Brassard, and A.~K. Ekert, Sci. Am. {\bf 267},  50  (1992).

\bibitem{Hughes95}
R. Hughes {\it et~al.}, Contemp. Physics {\bf 36},  149  (1995).

\bibitem{Unruh94}
W.~G. Unruh, Phys. Rev. A {\bf 51},  992  (1995).

\bibitem{Chuang95a}
I.~L. Chuang, R. Laflamme, P. Shor, and W.~H. Zurek, Science {\bf 270},  1633
  (1995).

\bibitem{Turchette95}
Q.~A. Turchette {\it et~al.}, Phys. Rev. Lett. {\bf 75},  4710  (1995).

\bibitem{Monroe95}
C. Monroe {\it et~al.}, Phys. Rev. Lett. {\bf 75},  4714  (1995).

\bibitem{Kraus83a}
K. Kraus, {\em States, Effects, and Operations} (Springer-Verlag, Berlin,
  1983).

\bibitem{Schumacher96a}
B.~W. Schumacher, LANL e-print quant-ph/9604023,  to appear in Phys. Rev. A
  (1996).

\bibitem{Louisell}
W.~H. Louisell, {\em Quantum Statistical Properties of Radiation} (Wiley, New
  York, 1973).

\bibitem{Gardiner91}
C.~W. Gardiner, {\em Quantum Noise} (Springer-Verlag, New York, 1991).

\bibitem{Mabuchi96}
H. Mabuchi, quant-ph/9608020  (1996).

\bibitem{Raymer94a}
M. Raymer, M. Beck, and D. McAlister, Phys. Rev. Lett. {\bf 72},  1137  (1994).

\bibitem{Leonhardt96}
U. Leonhardt, Phys. Rev. A {\bf 53},  2998  (1996).

\bibitem{Leibfried96a}
D. Leibfried {\it et~al.}, unpublished  (1996).

\bibitem{Horn91a}
R.~A. Horn and C.~R. Johnson, {\em Topics in matrix analysis} (Cambridge
  University Press, Cambridge, 1991).

\bibitem{Nielsen96c}
M.~A. Nielsen, B.~W. Schumacher, C.~M. Caves, and H. Barnum, in preparation
  (1996).

\bibitem{Lloyd96a}
S. Lloyd, LANL e-print quant-ph/9604015,  submitted to pra  (1996).

\bibitem{Schumacher96b}
B.~W. Schumacher and M.~A. Nielsen, LANL e-print quant-ph/9604022,  to appear
  in Phys. Rev. A  (1996).

\bibitem{braginsky92}
V.~B. Braginsky and F.~Y. Khalili, {\em Quantum Measurement} (Cambridge
  Unviersity Press, Cambridge, England, 1992).

\bibitem{Ljung87}
L. Ljung, {\em System Identification: Theory for the User} (Prentice Hall PTR,
  Upper Saddle River, 1987).

\end{thebibliography}


\end{multicols}

\end{document}